# Tuning many-body interactions in graphene: The effects of doping on excitons and carrier lifetimes

Kin Fai Mak[1]*, Felipe H. da Jornada[2,3]*, Keliang He[4], Jack Deslippe[2,3,5], Nicholas Petrone[6], James Hone[6], Jie Shan[4], Steven G. Louie[2,3], and Tony F. Heinz[1]

1. Departments of Physics and Electrical Engineering, Columbia University, New York, NY 10027, USA
2. Department of Physics, University of California, Berkeley, CA 94720, USA
3. Materials Sciences Division, Lawrence Berkeley National Laboratory, Berkeley, CA 94720, USA
4. Department of Physics, Case Western Reserve University, 10900 Euclid Avenue, Cleveland, OH 44106, USA
5. National Energy Research Scientific Computing Center, Lawrence Berkeley National Laboratory, Berkeley, CA 94720, USA
6. Department of Mechanical Engineering, Columbia University, New York, NY 10027, USA

*Contributed equally to this work

Abstract

The optical properties of graphene are strongly affected by electron-electron (*e-e*) and electron-hole (*e-h*) interactions. Here we tune these many-body interactions through varying the density of free charge carriers. Measurements from the infrared to the ultraviolet reveal significant changes in the optical conductivity of graphene for both electron and hole doping. The shift, broadening, and modification in shape of the saddle-point exciton resonance reflect strong screening of the many-body interactions by the carriers, as well as changes in quasi-particle lifetimes. *Ab initio* calculations by the GW Bethe-Salpeter equation (GW-BSE), which take into account modification of both the repulsive *e-e* and the attractive *e-h* interactions, provide excellent agreement with experiment. Understanding the optical properties and high-energy carrier dynamics of graphene over a wide range of doping is crucial for both fundamental graphene physics and for emerging applications of graphene in photonics.

Many of the distinctive properties of graphene, such as the characteristic linear dispersion relation of chiral massless Dirac fermions and the associated anomalous quantum Hall effect [1], can be understood within a picture of non-interacting electrons. These features are modified by many-body electronic interactions, leading to departures from the linear dispersion relation [2] and the presence of the fractional quantum Hall effect [3]. A similar situation prevails in terms of the optical response of graphene. The measured optical conductivity in the near-infrared and visible is close to the value of $e^2/4\hbar$ expected within the single-particle theory [4]. Significant departures from the predictions of single-particle theory are, however, revealed in the optical spectrum of graphene in the ultraviolet (UV) [5-9]. There the saddle point in the Brillouin zone at the M-point gives rise to a pronounced peak in the optical absorption. Both the position and shape of this feature reveal the role of strong Coulomb interactions. Instead of simple band-to-band transitions of free carriers, excitonic interactions give rise to resonances of correlated *e-h* pairs [5-12]. The same interactions also reduce the lifetime of high-energy quasiparticles [13-17].

The ability to tune the strength of the excitonic interactions and the decay rate of quasiparticles by varying the dielectric screening of the Coulomb interactions is highly important for both fundamental interests and an emerging area of graphene photonics [18]. But to date such a control of excitonic effects in the optical absorption of graphene by simple means (e.g. electrostatic gating) has not been demonstrated. In this paper, we demonstrate the possibility of tuning the many-body interactions in graphene by changing the density of free carriers. We probe the changes in the many-body interactions through their distinctive manifestation in the optical conductivity spectra of graphene. All significant aspects of the experimental results are reproduced by *ab-initio* calculations. In addition to the fundamental interest in the interactions of carriers in graphene, our findings show the possibility of modifying both the high-energy optical response of graphene and the decay processes of excited carriers. The latter suggests a means of tuning the Auger relaxation rates for more efficient carrier multiplication [19] and of controlling the response time of saturable absorption in graphene.

In our study, we measured the optical sheet conductivity $\sigma(E)$ of graphene over a broad range of photon energies $E$ from the near-infrared to the UV (1.2 ≤ $E$ ≤ 5.5 eV) for both electron and hole doping. Our experiments were performed using graphene field-effect transistors prepared on a transparent (quartz) substrate with a transparent electrolyte top gate [20]. Large-area graphene monolayers (~1 × 1 cm$^2$) were grown by chemical vapor deposition (CVD) on copper foil [21] and subsequently transferred to fused quartz substrates. Source and drain contacts to the graphene films were deposited by electron-beam evaporation. In order to probe doping levels as high as 10$^{14}$ cm$^{-2}$, we prepared transparent electrolyte top gates (PEO:LiClO$_4$ = 8:1), as described in Ref. [20]. Thin films of the transparent electrolyte were transferred to graphene samples and gold TEM grids (300 mesh) were installed as a contact to the electrolyte.

The optical conductivity of the graphene layer was established by absorption spectroscopy in transmission geometry. To access the desired spectral range of 1.2 to 5.5 eV, we made use of quartz tungsten halogen and deuterium light sources, respectively, for the visible and UV parts of the spectrum. The radiation was focused to a spot size ~ 500-µm in diameter on the sample. The transmission of the graphene sample was then measured using a grating spectrometer equipped with a UV-enhanced CCD-detector. For an accurate determination of the influence of doping, we recorded the fractional change in the transmission spectra, $1 - T(E)/T_0(E)$, of the graphene samples at different doping levels ($T(E)$) with respect to charge neutrality ($T_0(E)$). After taking into account the influence of the transparent substrate and top gate, these measurements yielded the doping-induced change in graphene absorbance $\Delta A(E)$ or, equivalently, in the graphene optical conductivity $\Delta\sigma(E)$. (See Supplementary Materials S1 for details.)

The measured optical conductivity spectrum $\sigma(E)$ of graphene displays marked changes as a function of carrier concentration (figure 1a). At low photon energies, we observe a progressive suppression of optical absorption with increasing doping. In addition, the peak in $\sigma(E)$ observed in the UV around 4.6 eV is found to be strongly influenced by the carrier density. The changes of the graphene response at low photon energies can be explained within a picture of state filling (Pauli blocking of the interband transitions) in the single-particle physics. The same is not true for the doping-induced changes in the UV response, since there is no modification of the occupation of the states involved in the optical transitions at the given doping levels. The UV response, associated with the saddle-point exciton resonance at the M-point of the graphene Brillouin zone [5-9], thus provides a direct signature of many-body interactions in graphene and of how carrier screening modifies these interactions.

We note two important trends in the UV response. For increased doping by either electrons or holes: (i) the peak of the saddle-point exciton resonance energy $\omega_0$ red shifts; and (ii) the peak feature becomes more symmetrical and broadens in width. The resonance energy $\omega_0$ shifts from 4.62 eV at charge neutrality to ~4.42 eV at high doping levels (figure 1b) [22], corresponding to an energy shift $\Delta\omega_0$ as large as 200 meV (figure 1c). In addition, up to experimental accuracy, we did not observe any increase in absorbance in the visible region upon doping, aside from the effect of Pauli blocking noted above.

To understand the experimental results, we need to evaluate the many-body effects in graphene and how the increased carrier screening in the doped samples modifies these effects. The observed changes in position and asymmetry of the saddle-point exciton feature with doping can be understood qualitatively on this basis. The emergence of a more symmetrical absorption feature with increased doping level reflects the reduced strength of *e-h* interactions due to enhanced carrier screening, since a symmetrical line shape is expected for band-to-band transitions at a 2D saddle-point singularity in the joint density of states [23]. The red shift in the peak position of the saddle-point exciton requires consideration of screening of both the attractive *e-h* interactions and the repulsive *e-e* interactions, with the latter dominating over the former. The data also reveal an increased width of the saddle-point exciton resonance with increasing doping. This effect suggests a decreased lifetime of these high-lying optically excited states in the more heavily doped system. This would reflect the emergence of additional decay channels, ones presumably associated with interactions of the highly-excited carriers with the free carriers near the Fermi surface. We have been able to substantiate all elements of this heuristic picture with rigorous *ab-initio* many-body theory, as well as to reproduce accurately the experimental absorption spectrum and its variation with doping level without the introduction of any adjustable parameters.

The theoretical approach is based on *ab initio* GW [24] and GW Bethe-Salpeter Equation (GW-BSE) calculations [25, 26] of the conductivity spectra at different doping levels (figure 1a), and the calculations were performed using the Berkeley GW package [27]. In our calculations, we used the DFT Kohn-Sham eigenvalues within the local density approximation (LDA) as our starting mean field for the GW and GW-BSE many-body calculations [24, 25, 28]. Our DFT-LDA calculations were performed on a 40x40x1 k-grid using a 60 Rydberg plane-wave cutoff for the electron orbitals and a norm-conserving Troullier-Martins pseudopotential. For the GW calculation, we chose a cutoff energy of 8 Ry for the dielectric matrix and evaluated the dielectric matrix $\varepsilon$ and the self-energy $\Sigma$ on a 80x80x1 k-grid. We employed the static remainder to accelerate convergence with respect to the number of unoccupied states. The absolute positions of the saddle-point peaks were converged to 20 meV and the relative peak shifts to 2 meV.

The self-energy was calculated for k-grid sizes up to 160x160x1, for which the total shifts were converged to 5 meV. The GPP model [27] was chosen to calculate the peak shifts and the $GW_0$ approach (*i.e.,* we self-consistently updated the Green's function) was used to capture the screening-dependent renormalization of the band-structure. For the BSE calculation, we evaluated the matrix elements on a 40x40x1 grid and interpolated onto a grid with 65,536 symmetry-inequivalent points. We employed the static approximation to the BSE. In order to evaluate the imaginary part of the self-energy $Im\,\Sigma$, the dielectric matrix was calculated with the same k-grid, and each of the frequencies was evaluated in an interval of 0.2 eV. The Green's function was also self-consistently updated in this case.

As part of the theoretical study, we also examined other possible effects associated with our experimental conditions. In particular, we investigated the possible influence of geometric relaxation of the graphene lattice structure with doping and the influence of the strong static electric fields associated with electrostatic doping. The former was found to have no meaningful effect on the predicted spectra, while the latter produced only a slight (< 25 meV) rigid blue shift in the UV spectrum. Both of these corrections have accordingly been neglected.

Figure 1a shows the absorption spectra calculated within the GW approximation (which do not include *e-h* interaction), which display a saddle-point resonance that is significantly blue-shifted from the experimental spectra. Moreover, the predicted red shift in this resonance with increasing doping is far stronger than that observed experimentally (figures 1a-c). On the other hand, when excitonic effects are included in the full GW-BSE calculations, the theory reproduces the experimental results quite accurately. The doping dependence of the peak position $\omega_0$ (figure 1b) and, particularly, of the peak shift $\Delta\omega_0$ (figure 1c) is in excellent agreement with experiment.

To understand the physics behind the shift $\Delta\omega_0$ of the saddle-point exciton energy with doping, we decompose this quantity into two contributions (figure 1c): (i) the change in the renormalization of quasiparticle self-energy, and (ii) the change in the excitonic correction to the spectrum. Both effects can be understood in terms of the influence of screening. This increased screening with increasing doping level reduces the energy needed for quasiparticles with a given momentum to be created, leading to the red shift associated with the first contribution. The enhanced screening also reduces the attractive *e-h* interactions. This reduces the original red shift from excitonic interactions in the neutral case. The excitonic contribution offsets about half of the red shift from decreased spacing of quasiparticle bands. The net result is a red shift of the saddle-point resonance $\Delta\omega_0$ with increasing doping that explains the experimental data quantitatively (figure 1c).

Although the GW-BSE calculations at this stage correctly capture the experimental peak shift with doping, the shape of the conductivity spectra calculated without lifetime effects of the quasiparticles significantly deviate from the observed spectral line shape (figure 1a). More specifically, the experimental excitonic features are all broader than those predicted by the calculations. While the calculations display sharper excitonic features at high doping levels than at low doping levels, the opposite trend is seen experimentally. We address this issue by including an *ab-initio* calculation of the excited-state lifetimes as a function of doping.

To determine the exciton lifetime, we consider separately the decay (*i.e.,* the imaginary part of the self-energy) of quasielectrons and quasiholes produced by optical excitation. We then derive the exciton lifetime by including the imaginary part of the quasiparticle energies perturbatively in the Bethe-Salpeter equation in a similar way as in Ref. [29]. The self-energy has contributions from *e-e* and electron-phonon (*e-ph*) interactions. We calculated the lifetimes arising from *e-e* interactions for four different doping levels and also included the effect of *e-ph* interactions from our previous study [17].

The optical conductivity of graphene calculated including lifetime effects in the GW-BSE calculations is in excellent agreement with experiment for both neutral and doped graphene (figure 2). The complete theory not only replicates the absolute exciton energy and the shift in energy with doping, as before, but also captures the change in width of the saddle-point exciton and its evolution to a more symmetric shape with increased doping. The agreement, we note, is obtained using a theory that contains no adjustable parameters.

It is instructive to compare the *ab-initio* theoretical predictions for the line width of the saddle-point exciton with the experimental broadening (figure 3). As expected from the agreement of the conductivity spectra, the experimental widths agree well with theory, both in magnitude and in their variation with doping density. For the case of theory, we further identify contributions to the broadening arising from *e-ph* and *e-e* interactions, respectively. (See supplementary materials S3 for details of extracting the broadening.)

The contribution to the exciton lifetime arising from *e-ph* interactions is seen to remain nearly constant as a function of doping. This behavior reflects the fact that near the saddle point, the imaginary part of the quasiparticle self-energy due to *e-ph* interactions, while varying with the quasiparticle energy, has little dependence on the doping level [17]. In contrast, the interactions of these highly-excited quasiparticles with other electrons, *i.e.,* the *e-e* interactions, increase with doping. We attribute this increase to the enlarged Fermi surface with doping, which increases the number of available decay channels for the quasiparticles [15, 17]. For these highly-excited quasiparticles, the decay from *e-e* interactions is not negligible even at low doping levels. We note that the decay process is influenced by the detailed band structure of graphene; the self-energy of quasiparticles for states near the saddle point is quite anisotropic and the broadening of the saddle-point resonance arises largely from the rapid decay of states in the $K \to \Gamma$ direction.

In addition to the changes in the quasiparticle decay rates, the line width of the optical absorption near the saddle-point resonance is also influenced by the range of free *e-h* pairs that participate in absorption at a specific photon energy [5]. This can be quantified by defining a k-space distribution of the oscillator strength from a narrow energy window of exciton states, $I(\mathbf{k}) = \sum_S |A_\mathbf{k}^S \langle v\mathbf{k}|\mathbf{v} \cdot \hat{e}|c\mathbf{k}\rangle|$ (figure 4a-b), where $A_\mathbf{k}^S$ is the coefficient of expansion of the exciton eigenstate in terms of the free quasiparticle states $|c\mathbf{k}\rangle$ and $v\mathbf{k}\rangle$ (Supplementary materials S3), $\langle v\mathbf{k}|\mathbf{v} \cdot \hat{e}|c\mathbf{k}\rangle$ is the independent particle optical transition matrix element, and the summation is over states S with energies $\Omega_S$ within a window of $\pm 75$ meV at the peak position. A similar distribution can be defined as a function of the quasiparticles' energies as $I(\Delta\omega) = \sum_{S,\mathbf{k}} |A_\mathbf{k}^S \langle v\mathbf{k}|\mathbf{v} \cdot \hat{e}|c\mathbf{k}\rangle| \delta[\Delta\omega - (\epsilon_{c\mathbf{k}} - \epsilon_{v\mathbf{k}} - \Omega_S)]$ (figure 4c-d). Upon increasing doping, both the k-space (figure 4b) and energy (figure 4d) distributions become more localized, which shows that there is less redistribution of oscillator strength among different regions of the band structure when the *e-h* interactions are screened. In fact, in the highly doped samples, the oscillator strength associated with the peak of the conductivity spectrum mainly comes from regions near the M point, which explains why the GW-BSE conductivity spectra approach the GW ones for highly doped samples.

The decrease in the number of participating *e-h* pair configurations upon doping is also important for the fine details of the line width of the saddle-point exciton. While the scattering rate of a state at a particular momentum tends to increase with doping, the reduction in the *e-h* correlation effects causes the overall line width of the exciton resonance to show a weak saturation in our theory (figure 3). The higher experimental values may be attributed to the influence of inhomogeneity in the doping level of the sample in experiment, as well as approximations in the theory from the omission of dynamical screening effects in the *e-h* interaction kernel [28] and substrate interactions [30].

In conclusion, our combined experimental and theoretical investigation has provided a coherent understanding of the nature of optical transitions in graphene and of the manner in which tuning the many-body interactions through doping alters the graphene optical conductivity spectrum. This knowledge is not only of fundamental importance, but also has significant implications for the rapidly developing areas of graphene photonics and optoelectronics.

The authors acknowledge helpful discussions with Dr. Mark Hybertsen. Support for the growth of the graphene samples was provided by the U.S. Department of Energy EFRC program (grant DESC00001085). Device fabrication was supported by the U.S. Office of Naval Research under the MURI program at Columbia University and by the National Science Foundation (grant DMR-0907477) at Case Western Reserve University. Optical characterization and analysis was supported by NSF (grant DMR-1106225). Theoretical formulation and study of lifetime effects was supported by the National Science Foundation (grant DMR10-1006184) and by the U.S. Office of Naval Research under the MURI program at University of California at Berkeley. The simulation and analysis of the optical absorption was supported by the Director, Office of Science, Office of Basic Energy Sciences, Materials Sciences and Engineering Division, U.S. Department of Energy under Contract No. DE-AC02-05CH11231. S. G. L. acknowledges support of a Simons Foundation Fellowship in Theoretical Physics. Computational calculations include the effects of screening of the quasiparticle self-energy, as reflected in a renormalized Fermi velocity.

**Figures and figure captions**

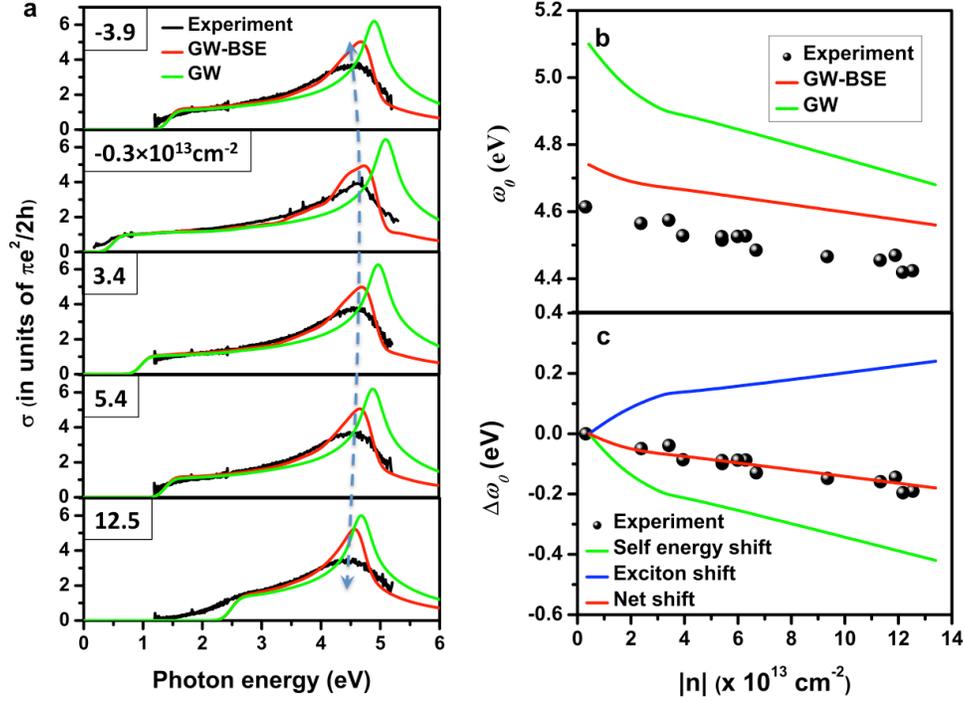

Fig. 1. Optical response of graphene as a function of doping, showing results of experiment (black), of GW calculations (green), and of GW-BSE calculations (red). (a) The sheet conductivity of graphene as a function of photon energy at different doping densities. A clear red shift of the experimentally observed peak in the UV is observed for both electron and hole doping. The position of the peak and its variation with doping are captured by GW-BSE calculations. The GW calculations, which omit *e-h* interactions, overestimate the resonance energy for all doping densities. (b) The resonance energy $\omega_0$ as a function of doping density, with combined data for electrons and holes. (c) The shift $\Delta\omega_0$ in resonance energy from that at charge neutrality as a function of doping, with combined data for electrons and holes. The net shift (red) calculated within the GW-BSE method is decomposed into two opposing contributions from screening of the electron self-energy (green) and of the excitonic response (blue). Effects due to finite lifetime of the excited electron and hole have not been included in the GW and GW-BSE calculations for the theoretical curves in this figure.

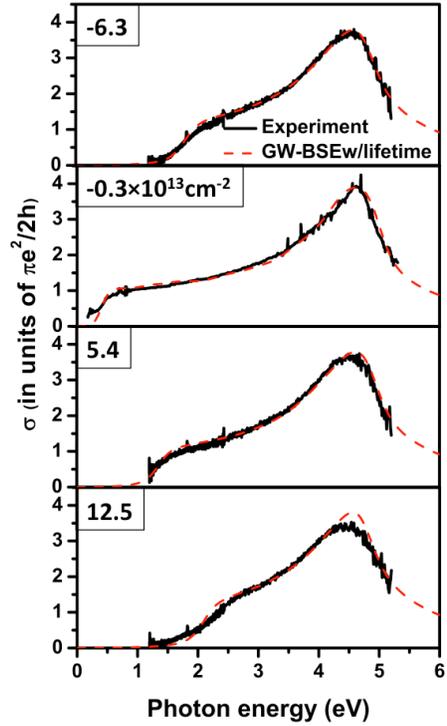

Fig. 2. *Ab initio* optical conductivity including lifetime effects. The calculated GW-BSE optical conductivity spectra with quasiparticle lifetime effects included (red dashed lines) are compared to the experimental spectra (black lines) for several doping densities. By evaluating the influence of both *e-e* and *e-ph* decay channels for the optically excited states, we obtain agreement with the experimental spectra for all doping levels.

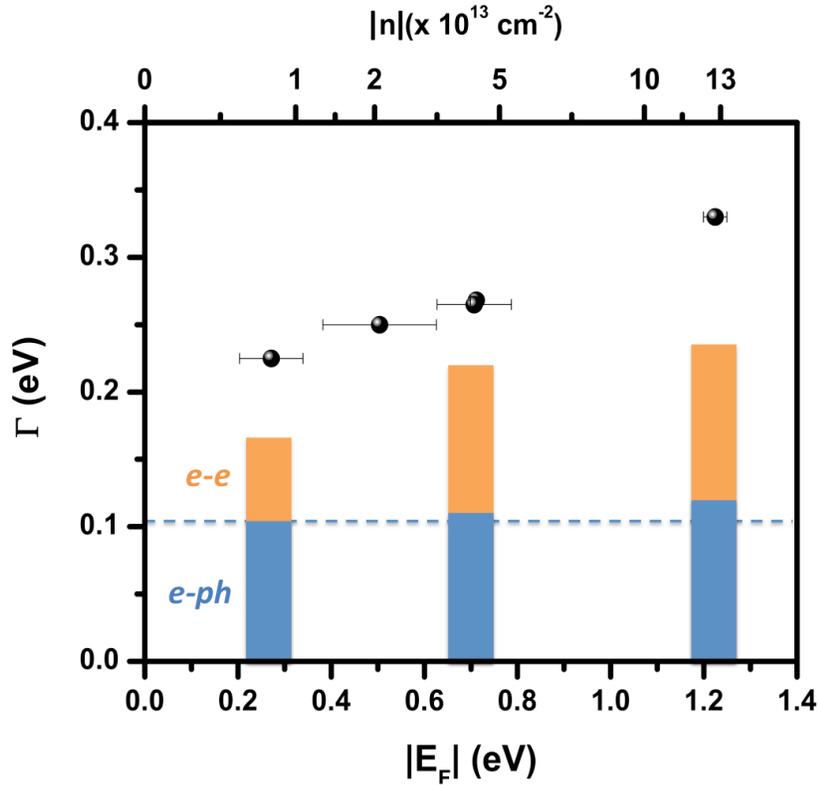

Fig. 3. Decay rates of highly excited quasiparticles in graphene. The line width of the saddle-point exciton (as discussed in the text) as a function of doping level, given in terms of the Fermi energy, for measurements with both electron and hole doping. The dots are the experimental values, while the columns show the calculated contributions from *e-ph* (blue) and *e-e* (orange) interactions. The horizontal doted line is a guide to the eye.

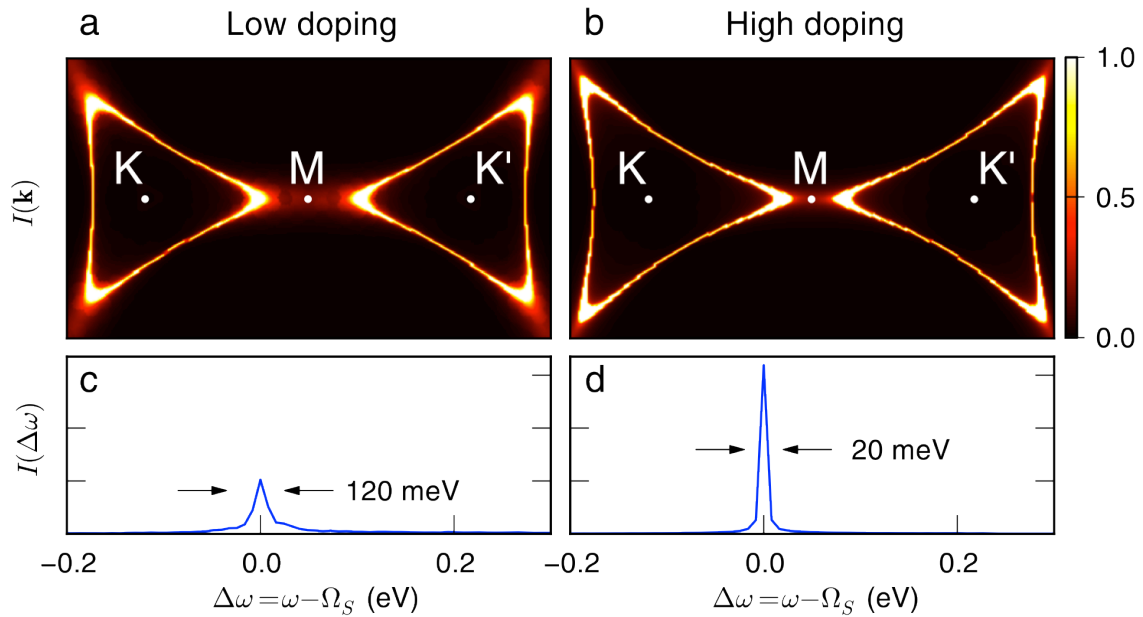

Fig. 4. Redistribution of oscillator strength. The k-space distribution of the oscillator strength $I(\mathbf{k})$ is shown for calculations with (a) a low carrier doping density ($n = 0.4 \times 10^{13}$ cm$^{-2}$) and (b) a high doping density ($n = 13 \times 10^{13}$ cm$^{-2}$). The corresponding energy distribution of oscillator strength $I(\Delta\omega)$ is also shown for (c) the low and (d) the high doping levels.

# Supplementary materials
# Tuning many-body interactions in graphene: The effects of doping on excitons and carrier lifetimes

Kin Fai Mak, Felipe H. da Jornada, Keliang He, Jack Deslippe, Nicholas Petrone, James Hone, Jie Shan, Steven G. Louie, and Tony F. Heinz

**Supplementary discussions**

**S1. Determination of the optical conductivity of graphene at different doping levels**

Consider a thin film of material with weak absorption located at the interface of two transparent materials. This describes the case of a layer of graphene placed upon a fused quartz substrate ($SiO_2$) and covered by a polymer electrolyte (PEO) gate, with corresponding refractive indices designated by $n_{SiO2}(E)$ and $n_{PEO}(E)$ as a function of photon energy $E$. The measured quantity in our experiments was the differential change in transmission, $T(E)$, from the transmission at neutrality $T_0(E)$, of the sample as a function of the doping level: $\delta = 1 - T(E)/T_0(E)$. By solving the Maxwell's equations, we can write (in cgs units) $\delta = \frac{2[A(E)-A_0(E)]}{[n_{SiO2}(E)+n_{PEO}(E)]} = \frac{8\pi}{c}\frac{[\sigma(E)-\sigma_0(E)]}{[n_{SiO2}(E)+n_{PEO}(E)]}$, where $A(E)$ ($\sigma(E)$) and $A_0(E)$ ($\sigma_0(E)$) are, respectively, the sample absorbance (real part of the sheet conductivity) for electrostatically doped and neutral graphene [1, 2]. The refractive index for amorphous $SiO_2$ is well characterized in the relevant spectral range [3]. To determine the refractive index of the polymer electrolyte, we have performed ellipsometry measurements (described below) in the corresponding spectral range. We can, therefore, directly convert the measured quantity $\delta$ into the differential change in sample absorbance and/or optical conductivity. Combing the absorbance and optical conductivity spectra for neutral graphene measured in ref. [2], we can then obtain the full spectrum of $A(E)$ and $\sigma(E)$ at different doping levels from our experimental data.

To perform the ellipsometry measurement on the polymer top gate, we spin-coated the polymer electrolyte (dissolved in methanol) on a fused quartz substrate (1-mm thickness from Chemglass) and dried the film at 110°C under ambient conditions. The optical properties of the resulting polymer film were measured using a spectroscopic ellipsometer (Horiba UVISEL) over the spectral range of $1.5 \leq E \leq 6.5$ eV. The ellipsometric parameters ψ and Δ (defined as $tan(\psi)exp(i\Delta) = r_p/r_s$, where $r_p$ and $r_s$ are the reflection coefficients for *p*- and *s*-polarized light components) were recorded at an angle of incidence of 57° with a 1-mm spot diameter. Measurements were taken for both the bare quartz substrate and the substrate covered by the polymer film; the results were modeled with DeltaPsi2 software for multilayer films [4]. We fit the spectra using a description that includes roughness in the polymer layer and dielectric functions for both the fused quartz substrate and polymer layer of an assumed Lorentzian form.

**S2. Determination of the doping density as a function of gate voltage from the measured differential conductivity spectra**

Figure S1a shows the change in optical conductivity Δσ(E) of graphene for different gate voltages. The behavior in the ultraviolet spectral range is our principal interest and has been discussed in the main text. However, for a determination of the carrier doping level, the relevant part of the spectrum lies in the visible ($E < 2$ eV). Here we see a progressive depletion of absorption at low photon energies as the doping level is increased by either electrons or holes. This behavior reflects the influence of Pauli blocking of transitions at photon energies $E$ below twice of the Fermi energy $E_F$ [5,6]. We use this spectroscopic signature to determine the absolute doping level n(V) as a function of gate voltage V (See main text). The results are shown in figure S1b. The gate capacitance at low applied voltages (based on the linear fit shown in the figure) is found to be 2.72 μFcm$^{-2}$. This value is more than 200 times greater than that obtained for a Si/SiO$_2$ back gate with the usual 300-nm oxide layer, in agreement with a recent study of highly doped graphene [7].

**S3. Determination of experimental and theoretical quasiparticle lifetimes**

As discussed in the main text, the predictions of the GW-BSE calculations shown in figure 1a provide a good match to the experimental determination of the shift in the peak of the optical absorption spectrum of graphene with doping density. However, without taking lifetime effects into account, the calculated spectra do not accurately reproduce both variation in the shape of the observed absorption feature and the broadening of the excitonic feature with increased doping. To determine the exciton lifetime, we consider separately the decay of quasielectrons and quasiholes produced by optical excitation, as discussed in the main text. Within this picture, we derive the following expression for the lifetime:

$$(\tau^S)^{-1} = \sum_{vc\bm{k}} \left| A^S_{vc\bm{k}} \right|^2 (\tau_{c\bm{k}}^{-1} + \tau_{v\bm{k}}^{-1}) ,$$

where $\tau^S$ is the exciton lifetime, $\tau_{n\bm{k}}$ is the quasiparticle lifetimes for band $n$ and wavevector $\bm{k}$, and $A^S_{vc\bm{k}}$ is the coefficient (from solving the Bethe-Salpeter equation) that defines the excitonic state $|S\rangle$ in terms of an expansion in free quasielectron and quasihole excitations $|S\rangle = \sum_{vc\bm{k}} A^S_{vc\bm{k}} |c,\bm{k}\rangle \otimes |v,\bm{k}\rangle$. Furthermore, the quasiparticle lifetime is directly related to the imaginary part of the self-energy through $2|\text{Im}\,\Sigma(\epsilon, \bm{k})| = \tau^{-1}$, where $\epsilon$ is the quasiparticle energy. A similar expression for $(\tau^S)^{-1}$ has also been derived when the quasiparticles' lifetimes arise solely from electron-phonon interactions [8].

Here we describe the method used to quantify with a single parameter the broadening of the spectra due to the lifetime of quasiparticles (figure 3). The procedure adopted was based on fitting the GW-BSE spectra obtained *without lifetime effects* both to the experimental data and to the theoretical results of the full calculation with quasiparticle decay. To this end, we matched the relevant (experimental or theoretical) spectrum to that generated from the GW-BSE spectrum under broadening by convolution with Lorentzian functions of appropriately chosen line widths. The results of such fit to the experimental spectra using the GW-BSE calculation that neglected lifetime effects are shown in figure S2. (For comparison in figure S2, we also display spectra broadened in the same manner, but based on the GW calculation that neglected lifetime effects.) This procedure for fitting the experimental results implicitly includes any influence of broadening arising from defects and sample inhomogeneity, factors not included in the calculated broadening shown in figure 2 of the main text.

## Supplementary figures

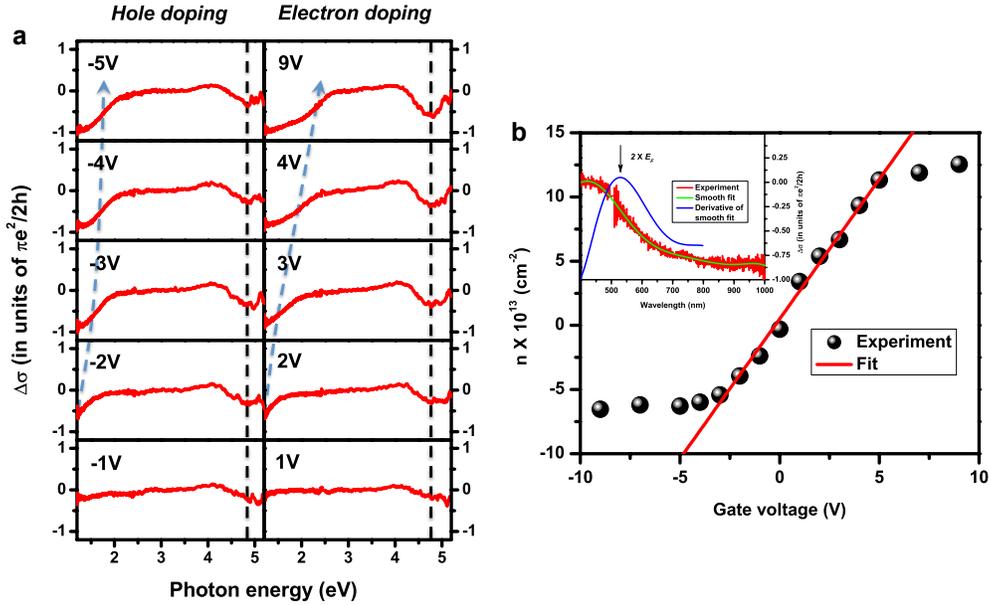

Fig. S1 (a) Measured change in the sheet conductivity of graphene, with respect to charge neutrality, from the near infrared to the ultraviolet for different gate voltages. The arrows trace the interband absorption threshold arising from Pauli blocking of interband optical transitions. The vertical dashed lines indicate, for reference, the energy where changes with doping are observed in the excitonic response in graphene. (b) Free-carrier density in graphene as a function of gate voltage as inferred from the spectra like those in (a). The linear fit yields an estimate of the gate capacitance for low gate voltages. The inset illustrates the procedure for the determination of the Fermi energy $E_F$ and doping density. We associate the value of $2E_F$ with the photon energy at which the steepest change in the absorption is found.

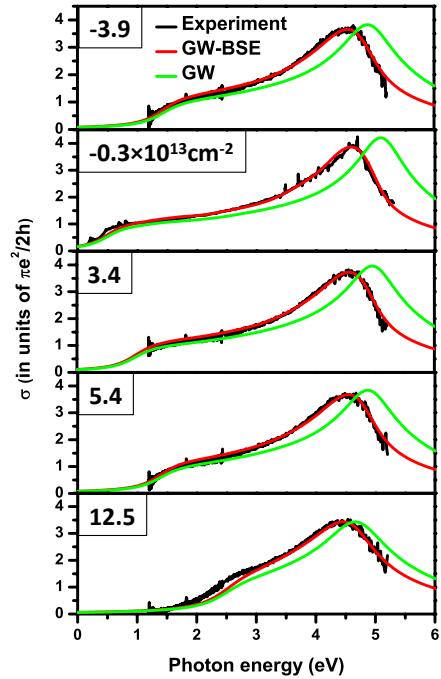

Fig. S2. The spectra of the sheet conductivity of graphene as a function of doping density. The figure shows spectra obtained from experiment (black) and from GW (green) and GW-BSE (red) calculations. The GW-BSE spectra here are obtained with GW-BSE calculated results that neglected quasiparticle lifetime effects but broadened by convolution with Lorentzians of appropriate line width to produce the best fit to the experimental results. The GW spectra have been broadened using the same parameters as for the GW-BS spectra.